An alternative to the topological interpretation of the transverse resistivity anomalies in $SrRuO_3$

Daisuke Kan[1,a)] Takahiro Moriyama[1, b)], Kento Kobayashi[1], and Yuichi Shimakawa[1,2]

[1]Institute for Chemical Research, Kyoto University, Uji, Kyoto 611-0011, Japan

[2]Integrated Research Consortium on Chemical Sciences, Uji, Kyoto 611-0011, Japan

electronic mail: [a)]dkan@scl.kyoto-u.ac.jp [b)] mtaka@scl.kyoto-u.ac.jp

We clarify the physical origin of anomalies in transverse resistivity often observed in exotic materials, such as $SrRuO_3$, in which the Berry curvature is manifested in the transport properties. The previously attributed mechanism for the anomalies, the topological Hall effect (THE) [e.g. *Sci. Adv.* 2, 160030 (2016)], is refuted by our thorough investigations as well as formulation of a model considering inhomogeneous magnetoelectric properties in the material. Our analyses fully explain every feature of the anomalies without resorting to the THE. The present results establish a fundamental understanding, which was previously overlooked, of magneto-transport properties in such exotic materials.

Materials and artificial heterostructures having a strong Dzyaloshinskii–Moriya (DM) interaction can host topologically nontrivial spin textures such as skyrmions[1,2]. Such topological magnetic textures can give rise to a Berry-curvature-originated fictitious magnetic field on electrons in motion and induce an additional transverse electron scattering known as the topological Hall effect (THE)[3-11]. While direct observation of nano-meter-scale skyrmions is experimentally challenging[12-15], the existence of the skyrmions is often inferred from the THE observed with a simple transport measurement. In fact, recent reports[16-19] have discussed that the anomalies in the transverse resistivity of various materials and heterostructures, including multilayers of $SrRuO_3$ (SRO), are attributed to the THE due to the formation of skyrmions.

The perovskite SRO has intriguing electric properties originating from the strong spin-orbit interaction[20] together with the multiple band crossings around the Fermi level. Due to the temperature-dependent band crossings and their Berry curvatures, the temperature dependence of the anomalous Hall resistivity $\rho_{AHE}$ does not simply follow that of the magnetization[21]. Most interestingly, $\rho_{AHE}$ can become zero at a certain temperature (which we call $T_S$.) while the magnetization is non-zero. SRO is therefore one of the rare materials in which one can identify an intrinsic anomalous Hall effect (AHE) originating from the Berry curvature[22].

In this study, we explore the transverse resistivity of single-layer films of SRO with different thicknesses ($t_{SRO}$) and epitaxially grown on $NdGaO_3$ (NGO) substrates. We particularly focus on SRO thickness $t_{SRO}$ = 3 – 4.5 nm, the range in which the variation of $T_S$ is significant. Every sample exhibits atypical humps, in the vicinity of $T_S$, in the transverse resistivity as a function of the applied magnetic field which resemble what is called the THE[16-19,25,26]. However, based on our thorough investigations including the thickness dependence of the appearance of the humps, minor loop measurements of the transverse resistivity, and a numerical modeling, we discuss an alternative and more plausible mechanism explaining these anomalies.

We epitaxially grew SRO thin films on (110) NGO substrates by pulsed laser deposition. The SRO layer was deposited by pulsing $SrRu_{1.3}O_x$ targets with a KrF excimer laser ($\lambda = 248$ nm). We confirmed by x-ray diffraction measurements that the (110)-oriented SRO layer was coherently grown on the substrates (~1.7% compressive strain). Very smooth step-and-terrace surface structures with single pseudo-cubic unit cell height steps (~4Å) were observed by atomic force microscopy. Longitudinal and transverse electrical resistivities ($\rho_{xx}$ and $\rho_{xy}$) were measured by a conventional Van der Pauw method.

Figure 1 shows temperature dependence of $\rho_{xx}$ for $t_{SRO} = 3 – 4.5$ nm. For all the samples, $\rho_{xx}$ overall decreases with decreasing temperature, indicating a metallic conduction. The ferromagnetic transition can be identified by the slight change in each curve in Fig. 1. We define the transition temperature $T_C$ as the temperature at which the AHE vanishes. (See Fig. 2, for example). It is found that $T_C$ slightly decreases with decreasing $t_{SRO}$.

In contrast to the thickness dependence of $\rho_{xx}$, a slight difference in $t_{SRO}$ significantly impacts on the behavior of $\rho_{xy}$. Figure 2 shows magnetic field $H$ dependence of $\rho_{xy}$ for $t_{SRO} = 3.5$ and 4.5 nm (the data set for $t_{SRO} = 3$ and 4 nm is provided in the supplementary information (SI)). Note that the component of the ordinary Hall effect was subtracted from all the data shown in Fig. 2. The square hysteresis loop of $\rho_{xy}$-$H$ plots reflects the AHE in response to the magnetization switching. It is clear that there are two intriguing features in the hysteresis loops. One is that the squareness of the hysteresis loop as well as the polarity of $\rho_{xy}$ varies with temperature. The other is that some atypical humps around the magnetization switching field are observed in a certain temperature range. Here we define the anomalous Hall resistivity $\rho_{AHE}$ as the saturation resistivity in the positive field and also define $\rho_{hump}$ as the height of the hump with respect to the saturation resistivity and $H_{\rho\_peak}$ as the field at which the hump is positioned. These definitions are indicated in Fig. 2. ($\rho_{AHE}$ can be either positive or negative, depending on the polarity of the hysteresis loop.).

The sample with $t_{SRO}$ = 3.5 nm ($T_C \sim 120$ K), for instance, shows a positive $\rho_{AHE}$ at high temperature and undergoes reversal of the sign of $\rho_{AHE}$ at $T_S$ = 68 K, where $T_S$ is defined as the temperature at which the sign of $\rho_{AHE}$ reverses. Films with other thicknesses essentially show similar trends with different $T_S$ values. The temperature dependence of $\rho_{AHE}$ is consistent with the previous reports and originates from the temperature-induced changes in the integrated Berry curvature over the electron distributions around the Fermi level[21,22]. The humps seen in $\rho_{xy}$-$H$ curves look quite similar to what were observed in previous reports[16,19,25,26] and were claimed to be due to the emergence of the THE. It should be emphasized that anomalies, essentially same as our observed humps, have been reported in SrIrO$_3$/SRO heterostructures[16,19], which are considered to host skyrmions due to a strong interfacial DM interaction. One may conceive that a structural asymmetry owing to the Ru-O-Ru bond angle variations [23,24] in the present sample can give rise to the DM interaction and form skyrmions, and therefore the THE could be present. In the following, however, we refute the THE mechanism and discuss an alternative physical origin for the appearance of the humps.

Figure 3 summarizes the temperature dependence of $\rho_{hump}$, $H_{\rho\_hump}$, $\rho_{AHE}$, and $H_C$ for $t_{SRO}$ = 3 – 4.5 nm. $\rho_{hump}$ is found to be always positive regardless of the film thickness. The maximum value of the $\rho_{hump}$ increases with decreasing $t_{SRO}$ and the temperature range where $\rho_{hump}$ is seen also becomes wider as $t_{SRO}$ decreases. It is found that $\rho_{hump}$ is maximized at $T_S$, and concomitantly the $H_c$ exhibits a discontinuity while $H_{\rho\_hump}$ smoothly changes across $T_S$. These behaviors of $\rho_{hump}$ and $H_c$ become more prominent for thinner films. We note that similar temperature dependences of $\rho_{hump}$ are seen for the tensilely strained SrRuO$_3$ films on GdScO$_3$ (GSO) substrates[27]. Given that the types of the substrate-induced strain (either compressive or tensile) and spatial dependence of the Ru-O-Ru bond angle across the interface differ between the films on NGO and GSO, structurally induced properties of general interest, such as DM interaction, would be irrelevant to the emergence of the humps seen in $\rho_{xy}$-$H$ plots.

We also investigated minor loops of $\rho_{xy}$-$H$ for $t_{SRO}$ = 3 nm at 20, 35 and 50K. The results are summarized in Figure 4. At each temperature, the loop starts from the positive field toward the negative field around which the hump appears and is folded back to the initial positive field. The maximum negative field in the minor loop measurements is referred to as $H_{n\_max}$. We essentially found that the humps are hysteretic, meaning that the appearance of them depends on how the minor loop is scanned. For instance, looking at the loops at 20 K, where the humps are the most significant, one can see that the emergence of the humps in the positive field seems to depend on whether or not $H_{n\_max}$ surpasses magnetic fields in which the hump is seen. One clearly sees in the $H_{n\_max}$ dependence of the $\rho_{hump}$ in the positive field shown in Fig. 4d that the $\rho_{hump}$ decreases to zero when $|H_{n\_max}| < |H_{\rho\_hump}|$.

We point out that the $H_{n\_max}$-dependent appearance of the humps cannot be in line with the story of the skyrmion formations leading to the THE unless one makes a rather convenient assumption that magnetic hysteresis in the skyrmions and the domains behaves as such[18]. Instead, we explain our overall experimental observations on the atypical humps by using a traditional magnetism taking into account the fact that $\rho_{AHE}$ and $H_C$ are strongly temperature dependent (see Fig. 3) and hypothesizing that they are inhomogeneous over the SRO film. We show in the following that those peculiar humps are indeed well reproduced by our model without considering the THE.

By starting with a simple toy model shown in the SI, main features of the humps can already be reproduced by considering the two domains (domain A and B) that contain different $T_S$ ($T_{S\_A}$ and $T_{S\_B}$, respectively). Here we show a complete reproduction of the hysteresis loops by more rigorous model taking into account multiple domains with a distribution of $T_S$ denoted by $T_\sigma$.

Considering the temperature dependence of $\rho_{AHE}(T)$ and $H_c(T)$, one can write a field response of the transverse resistivity in each domain having a given effective temperature $T'$ as

$$f(T') = \rho_{AHE}(T')\{1 - 2\mathrm{H}_{\mathrm{Heav}}(H - H_c(T'))\}g(T'), \tag{1}$$

where $\mathrm{H}_{\mathrm{Heav}}(x)$ is the Heaviside step function describing the magnetization switching and $H$ is the applied field. $g(T, T_\sigma)$ is the Gaussian function taking a distribution of the domain as

$$g(T') = \frac{1}{\sqrt{2\pi T_\sigma^2}} \exp\left(-\frac{(T'-T)^2}{2T_\sigma^2}\right). \tag{2}$$

Note that we implicitly assume that $\rho_{AHE}$ and $H_c$ are linear to $T$ so that the actual spatial distribution of $\rho_{AHE}$ and $H_c$ can be mapped as a function of the effective temperature as $\rho_{AHE}(T')$ and $H_c(T')$ (See Fig. S3 in SI for more detail.). For our calculation, $\rho_{AHE}(T')$ and $H_c(T')$ are taken from the actual temperature dependence of $\rho_{\mathrm{AHE}}$ and $H_c$ shown in Fig. 3 so that the only unknown parameter becomes $T_\sigma$

The $\rho_{\mathrm{AHE}}$-$H$ plot at a measurement temperature $T$ can be obtained by integrating $f(T', H)$ over $T'$,

$$\Gamma(H) = \int_0^\infty f(T')dT' \tag{3}$$

Note that, as $\Gamma(H)$ describes a magnetization switching in only one direction of the field sweep, the full loop is produced by taking another $\Gamma(H)$ for the other field sweep direction. Figure 5a shows the $\rho_{AHE}$-$H$ loops calculated with $T_\sigma$ = 10.7 K for the 3.5-nm-thick film. The hysteresis loops at various temperatures around $T_s$ ($T_s$ = 68 K for the 3.5nm-thick film) reproduce very well our experimental observations (See Fig. 2b). We also show in Fig. 5b that the temperature dependence of the $\rho_{\mathrm{hump}}$ extracted from the calculated loops (Fig. 5a) reproduces the experimentally obtained temperature dependence shown in Fig. 3. In particular, $\rho_{hump}$ is found to be maximized at $T_s$ and this behavior is exactly what is experimentally observed in the temperature dependence of the $\rho_{\mathrm{hump}}$. Our model highlights that a spatial variation of $T_s$ in the film essentially gives rise to a mixture of hysteresis loops with both positive and negative $\rho_{\mathrm{AHE}}$ around $T_s$, consequently leading to the emergence of the humps which is totally irrelevant to THE or skyrmion formation.

Our model coherently explains the experimentally observed temperature-dependent $\rho_{\mathrm{hump}}$

for SRO films having other thicknesses (not shown). It is found that reducing $t_{SRO}$ not only lowers $T_S$ but also increases the inhomogeneity of $T_S$ characterized by an increase of $T_\sigma$ (See SI for the estimation of $T_\sigma$). We also note that the minor loops of $\rho_{xy}$–$H$ (Fig. 4) can also be reproduced well by our model. Representative loops are shown in Fig. 5c, which clearly demonstrates that the humps in the positive field appear only when $|H_{n\_max}|$ is greater than $|H_{\rho\_hump}|$.

In summary, we showed that a single layer of SRO epitaxially grown on NGO substrates exhibits atypical humps in the transverse resistivity as a function of the external field, which resembles what has been claimed to be the topological Hall effect. However, our thorough investigations including the $t_{SRO}$ dependence of the appearance of the humps, minor loop measurements, and numerical modeling indicate that the topological Hall effect cannot be the only origin of the observed humps. Our model, assuming a spatial variation of $T_S$ in the film, reproduced every feature in the transverse resistivity very well, which strongly indicates that film inhomogeneities are the key factor responsible for the atypical humps. Our analysis further revealed that the variation of $T_S$ as small as 10.7 K is enough to replicate the humps. We would like to emphasize that, based on our model, these atypical humps in the transverse resistivity could be observed in other materials, for example, rare earth–transition metal alloys[28,29], if $T_S$ and $H_c$ are spatially varied within a film. Finally, the present results provide a fundamental understanding of magneto-transport properties in such exotic materials, which would impact recently flourishing studies on topological materials.

This work was partially supported by a grant for the Integrated Research Consortium on Chemical Sciences, by Grants-in-Aid for Scientific Research (Nos. 16H02266, 17H05217, 17H04924, 17H04813, by a JSPS Core-to-Core program (A), and by a grant for the Joint Project of Chemical Synthesis Core Research Institutions from the Ministry of Education, Culture, Sports,

References

[1] S. Mühlbauer, B. Binz, F. Jonietz, C. Pfleiderer, A. Rosch, A. Neubauer, R. Georgii, and P. Böni, Science **323**, 915 (2009).

[2] X. Z. Yu, Y. Onose, N. Kanazawa, J. H. Park, J. H. Han, Y. Matsui, N. Nagaosa, and Y. Tokura, Nature **465**, 901 (2010).

[3] M. Uchida and M. Kawasaki, J. Phys. D: Appl. Phys. **51**, 143001 (2018).

[4] S. Chakraverty *et al.*, Phys. Rev. B **88**, 220405 (2013).

[5] A. Fert, N. Reyren, and V. Cros, **2**, 17031 (2017).

[6] C. Moreau Luchaire *et al.*, Nat. Nano. **11**, 444 (2016).

[7] I. Kezsmarki *et al.*, Nat. Mater. **14**, 1116 (2015).

[8] Y. Ohuchi, Y. Kozuka, M. Uchida, K. Ueno, A. Tsukazaki, and M. Kawasaki, Phys. Rev. B **91**, 245115 (2015).

[9] N. Nagaosa and Y. Tokura, Nat. Nano. **8**, 899 (2013).

[10] M. Lee, W. Kang, Y. Onose, Y. Tokura, and N. P. Ong, Phys. Rev. Lett. **102**, 186601 (2009).

[11] A. Neubauer, C. Pfleiderer, B. Binz, A. Rosch, R. Ritz, P. G. Niklowitz, and P. Böni, Phys. Rev. Lett. **102**, 186602 (2009).

[12] A. Soumyanarayanan *et al.*, Nat. Mater. **16**, 898 (2017).

[13] S. Woo *et al.*, Nat. Mater. **15**, 501 (2016).

[14] S. Woo *et al.*, Nature Communications **9**, 959 (2018).

[15] W. Jiang *et al.*, Nat Phys **13**, 162 (2017).

[16] J. Matsuno, N. Ogawa, K. Yasuda, F. Kagawa, W. Koshibae, N. Nagaosa, Y. Tokura, and M. Kawasaki, Science Advances **2**, :e160030 (2016).

[17] K. Yasuda *et al.*, Nat Phys **12**, 555 (2016).

[18] B. M. Ludbrook, G. Dubuis, A. H. Puichaud, B. J. Ruck, and S. Granville, Scientific reports **7**, 13620 (2017).

[19] Y. Ohuchi, J. Matsuno, N. Ogawa, Y. Kozuka, M. Uchida, Y. Tokura, and M. Kawasaki, Nature Communications **9**, 213 (2018).

[20] G. Koster, L. Klein, W. Siemons, G. Rijnders, J. S. Dodge, C.-B. Eom, D. H. A. Blank, and M. R. Beasley, Reviews of Modern Physics **84**, 253 (2012).

[21] Z. Fang *et al.*, Science **302**, 92 (2003).

[22] N. Nagaosa, J. Sinova, S. Onoda, A. H. MacDonald, and N. P. Ong, Reviews of Modern Physics **82**, 1539 (2010).

[23] R. Aso, D. Kan, Y. Shimakawa, and H. Kurata, Scientific reports **3**, 2214 (2013).

[24] R. Aso, D. Kan, Y. Fujiyoshi, Y. Shimakawa, and H. Kurata, Crystal Growth & Design **14**, 6478 (2014).

[25] I. Lindfors-Vrejoiu and M. Ziese, physica status solidi (b) **254**, 1600556 (2017).

[26] B. Pang, L. Zhang, Y. B. Chen, J. Zhou, S. Yao, S. Zhang, and Y. Chen, ACS Applied Materials & Interfaces **9**, 3201 (2017).

[27] D. Kan and Y. Shimakawa, physica status solidi (b), in press (2018).

[28] T. Okuno, K.-J. Kim, T. Tono, S. Kim, T. Moriyama, H. Yoshikawa, A. Tsukamoto, and T. Ono, Applied Physics Express **9**, 073001 (2016).

[29] J. Finley and L. Liu, Physical Review Applied **6**, 054001 (2016).

Figure Captions

Figure 1: Temperature dependence of $\rho_{xx}$ for SRO films 3, 3.5, 4, and 4.5 nm thick.

Figure 2: Magnetic field dependence of $\rho_{xy}$ for (a) 4.5- and (b) 3.5-nm-thick SRO films, exhibiting atypical humps around the magnetization switching field. The $\rho_{xy}$-$H$ loops in the figures were obtained at various temperatures below the ferromagnetic transition temperature $T_C$ (140K for a 4.5-nm-thick film and 120K for a 3.5-nm-thick film). Every loop in the figure has an offset of 0.4 μΩcm.

Figure 3: $\rho_{AHE}$, $\rho_{hump}$, $H_{\rho\_hump}$, and $H_c$ as a function of temperature for SRO films (a) 3, (b) 3.5, (c) 4, and (d) 4.5 nm thick.

Figure 4: Minor loops of $\rho_{xy}$ for the 3-nm-thick film, revealing the $H_{n\_max}$-dependent appearance of the humps that cannot be in line with the story of the skyrmion formations leading to the THE. The loops were measured at (a) 20 K, (b) 35 K and (c) 50 K. (d) $\rho_{hump}$ and $H_{\rho\_hump}$ as a function of the maximum negative magnetic field $H_{n\_max}$.

Figure 5: (a) $\rho_{AHE}$-$H$ hysteresis loops reproduced by our numerical model with $T_\sigma = 10.7$ K, highlighting that film inhomogeneities are the key for the atypical humps in $\rho_{AHE}$. For the calculations, $\rho_{AHE}$ and $H_c$ experimentally observed for the $t_{SRO} = 3.5$ nm film were used. (b) Temperature dependence of the calculated $\rho_{hump}$. (c) Reproduced minor loops at $T = T_S$.

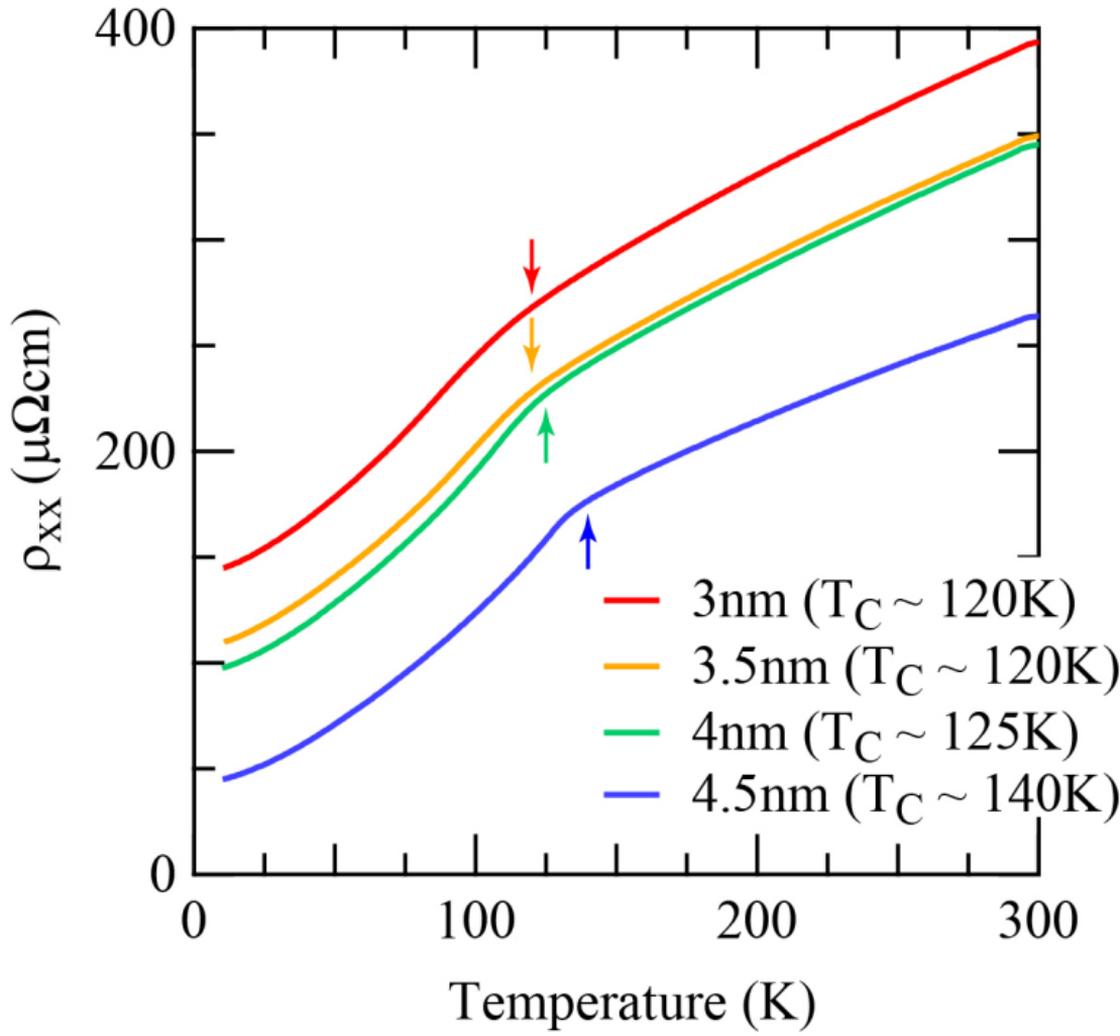

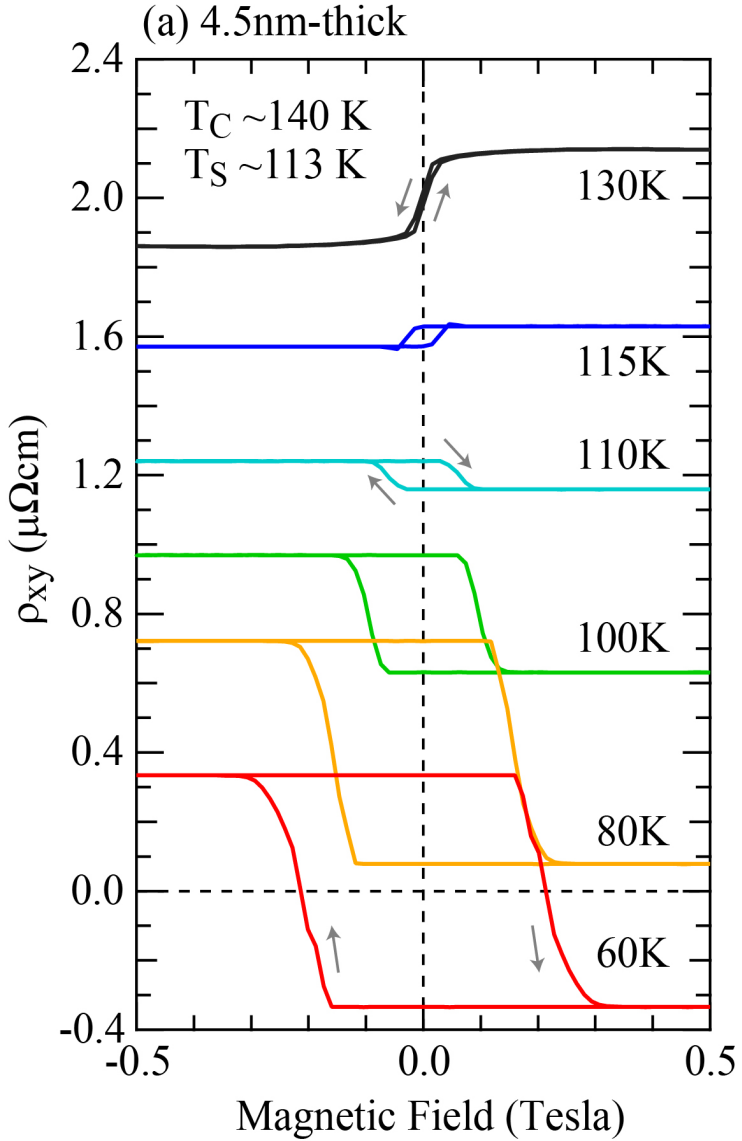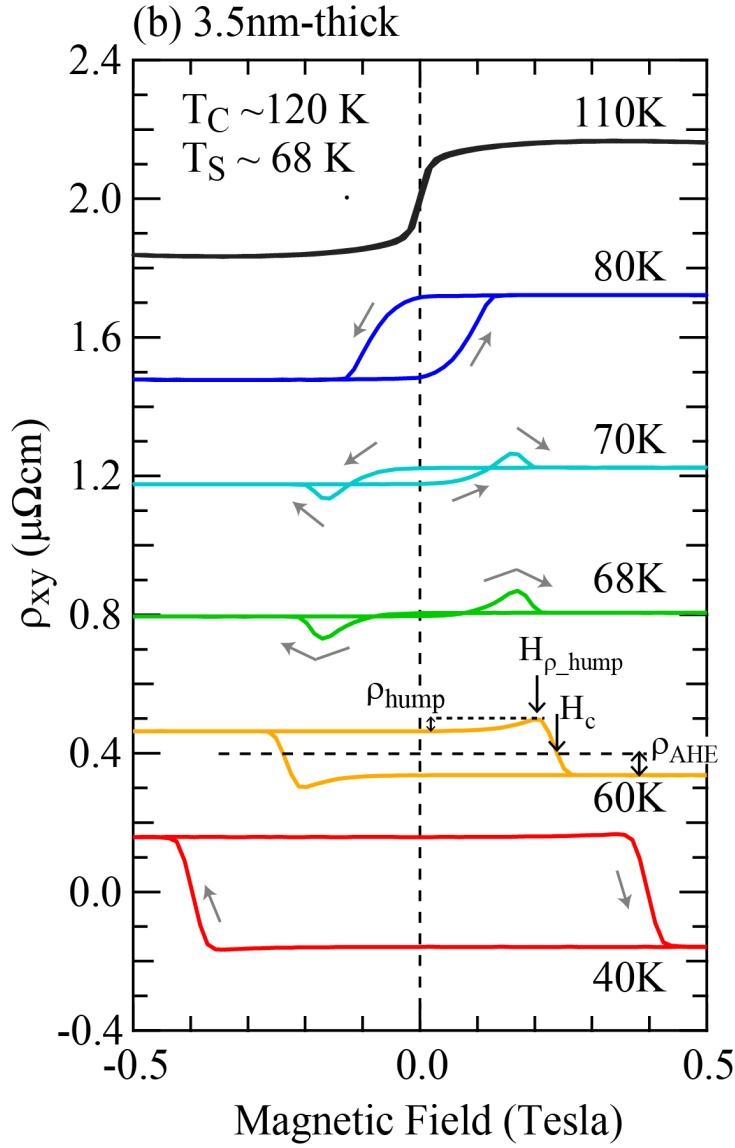

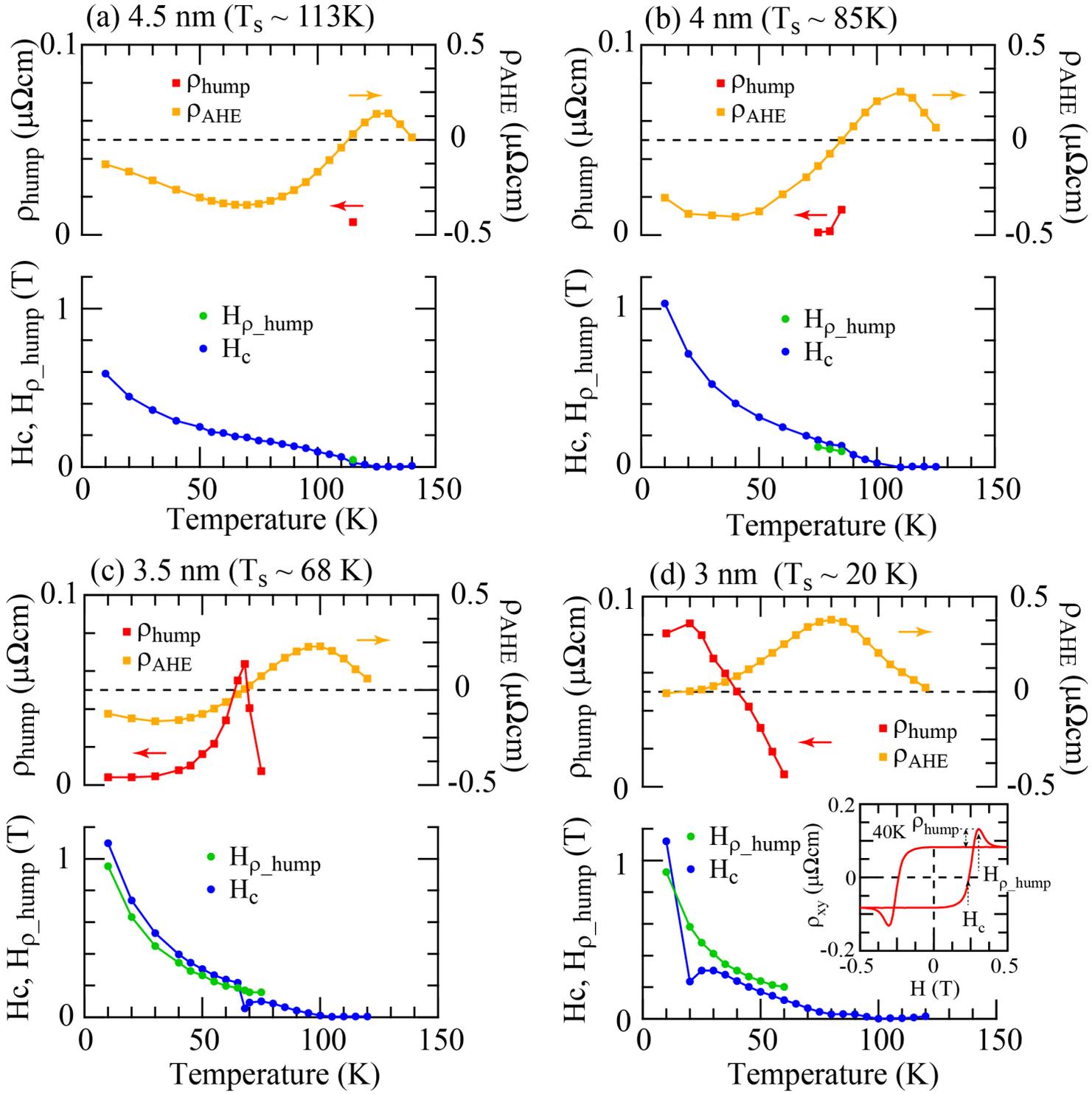

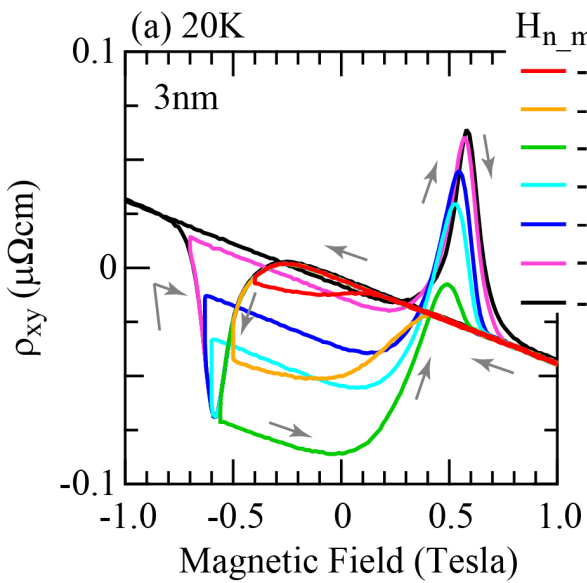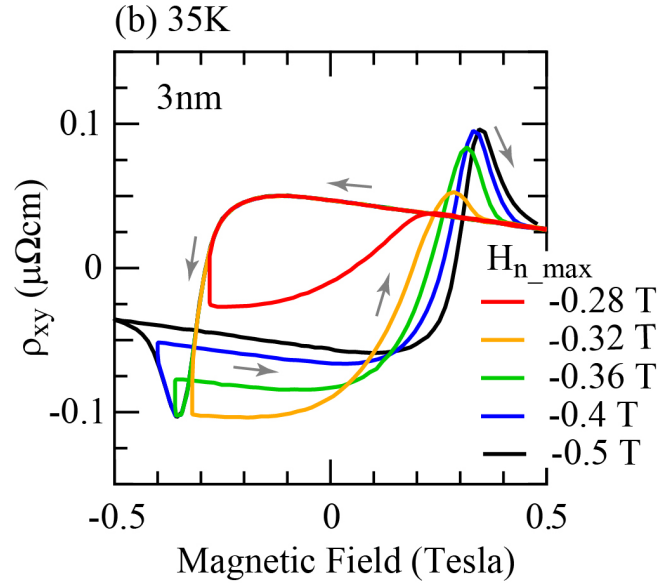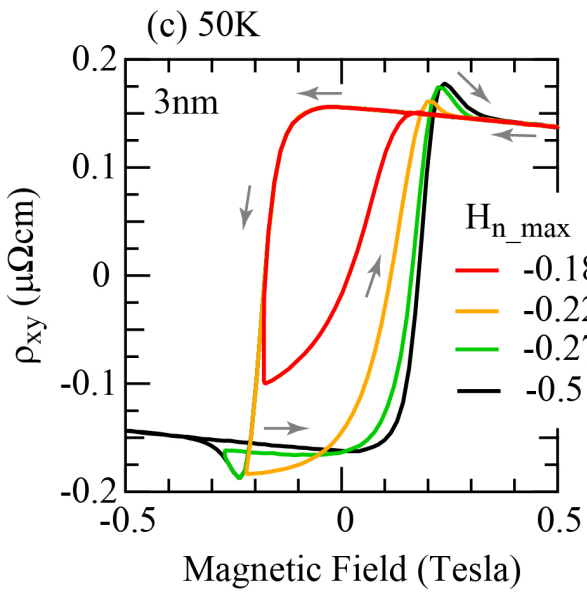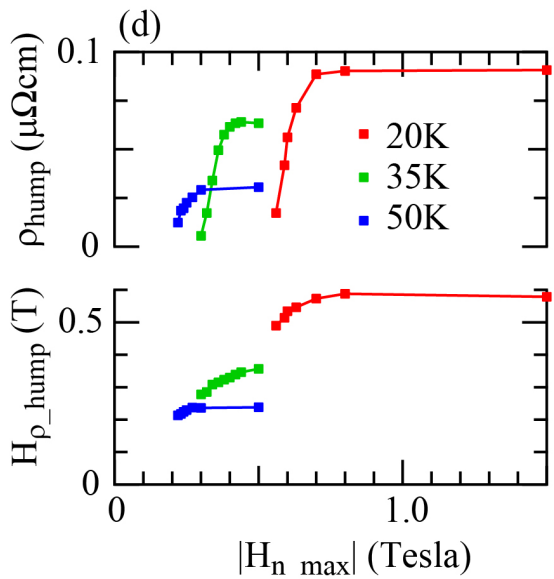

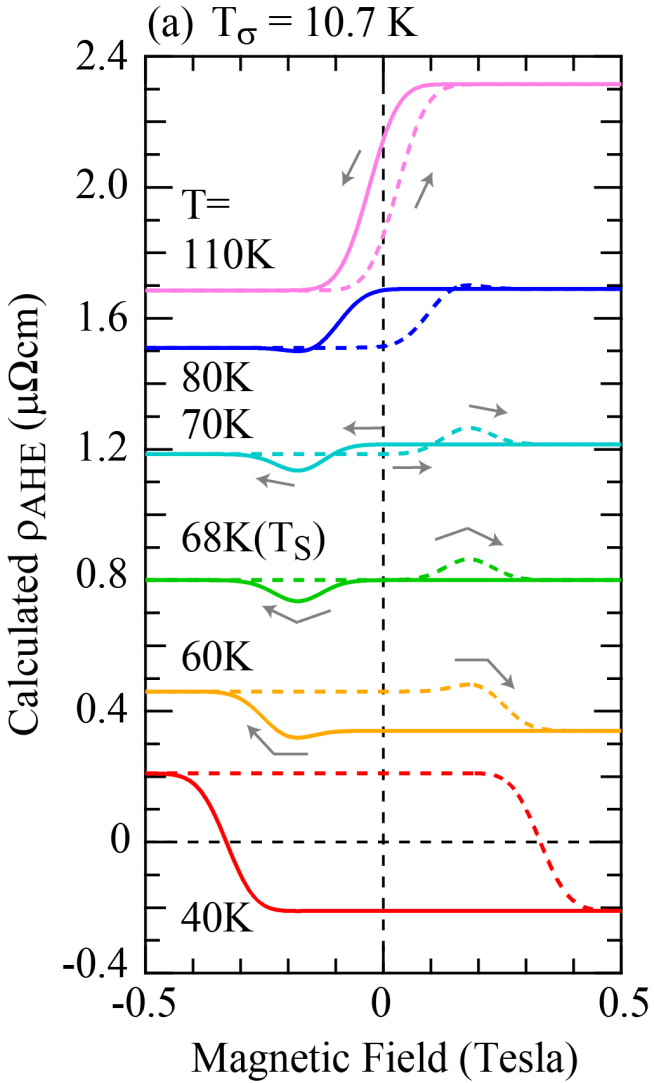
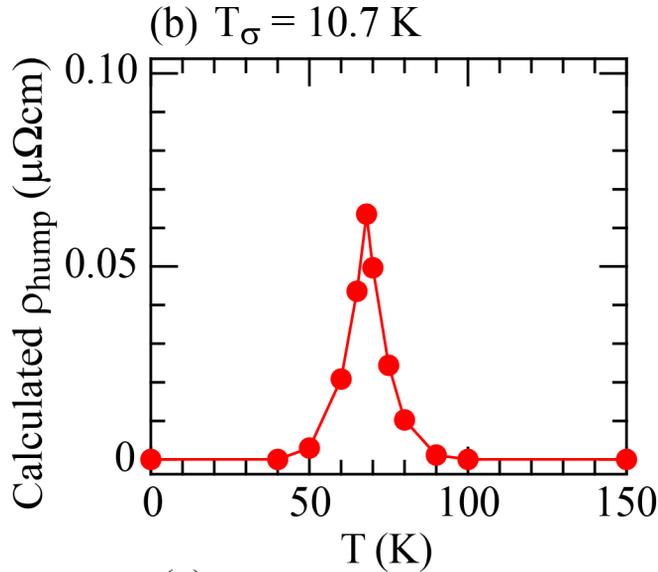
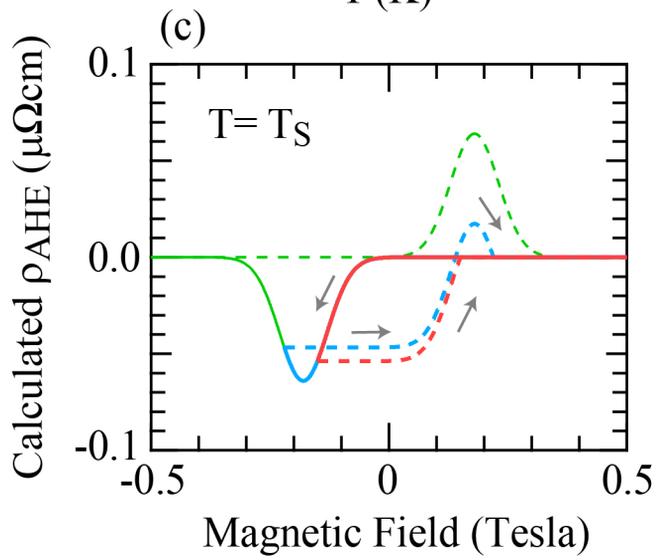